# Rifts in Spreading Wax Layers

Rolf Ragnarsson, J. Lewis Ford[*], Christian D. Santangelo and Eberhard Bodenschatz[**]
*Laboratory of Atomic and Solid State Physics, Cornell University, Ithaca, New York 14853-2501*
(version 10-18-95)

We report experimental results on the rift formation between two freezing wax plates. The plates were pulled apart with constant velocity, while floating on the melt, in a way akin to the tectonic plates of the earth's crust. At slow spreading rates, a rift, initially perpendicular to the spreading direction, was found to be stable, while above a critical spreading rate a "spiky" rift with fracture zones almost parallel to the spreading direction developed. At yet higher spreading rates a second transition from the spiky rift to a zig-zag pattern occurred. In this regime the rift can be characterized by a single angle which was found to be dependent on the spreading rate. We show that the oblique spreading angles agree with a simple geometrical model. The coarsening of the zig-zag pattern over time and the three-dimensional structure of the solidified crust are also discussed.

During the past two decades experiments with paraffin wax were used to model fundamental geophysical processes of the earth's crust [1-4]. One problem addressed was the structure of the mid-ocean ridges as observed between two separating tectonic plates [1-3]. In the corresponding experiment molten wax was frozen at the surface by a flow of cold air. Then the solid crust was pulled apart with constant velocity and a rift was formed separating the crust into two solid plates. A straight rift, initially perpendicular to the pulling direction, was found to evolve into a pattern consisting of straight segments interrupted by faults parallel to the pulling direction. The pattern was interpreted to resemble the transform faults of the mid-ocean ridges and a model based on shrinkage and the mechanical properties of the wax was proposed [2]. At slower spreading rates a rift consisting of oblique segments was also observed [2-3]. However, no further investigation of this latter phenomenon was undertaken and to date any detailed knowledge of the phase diagram is lacking.

In this Letter we report on the systematic experimental investigation of the rift formation as a function of spreading velocity. We found several distinct spreading regimes, each characterized by a unique type of rift pattern. At low spreading velocities we discovered a novel, "spiky" rift pattern dominated by fracture zones. At higher pulling speeds we found a transition to a regular "zig-zag" pattern, which with increasing pulling speed became steeper and steeper, finally forming a pattern of fault-like slips parallel to the spreading direction separated by straight regions perpendicular to it. This latter pattern is quite similar to the one observed and interpreted earlier as transform faults [1-3, 5]. We show that in the zig-zag regime the angle of the oblique rift segments can be described by a simple geometrical model. This new mechanism connects naturally to the transform fault regime, suggesting a pattern formation mechanism also found in other front propagation problems [6].

In the experiment, Shellwax 120, a macrocrystalline paraffin wax [7], was used. As shown in earlier experiments [2] it is well suited for the investigation of transform fault patterns. Shellwax 120 crystallizes at $54°C$ and has a solid-solid phase transition at about $35°C$ [8]. While the first solid phase is plastic and easily deformable, representing the earth's asthenosphere, the second is hard and brittle representing the earth's crust. Upon freezing, the wax shrinks in the temperature range from $54°C$ to $46°C$ by roughly 10% and by another 4% from $35°C$ to $28°C$ [8].

The experimental setup, as shown schematically in Fig. 1, consists of a rectangular tray of dimensions 114x36x10 cm$^3$

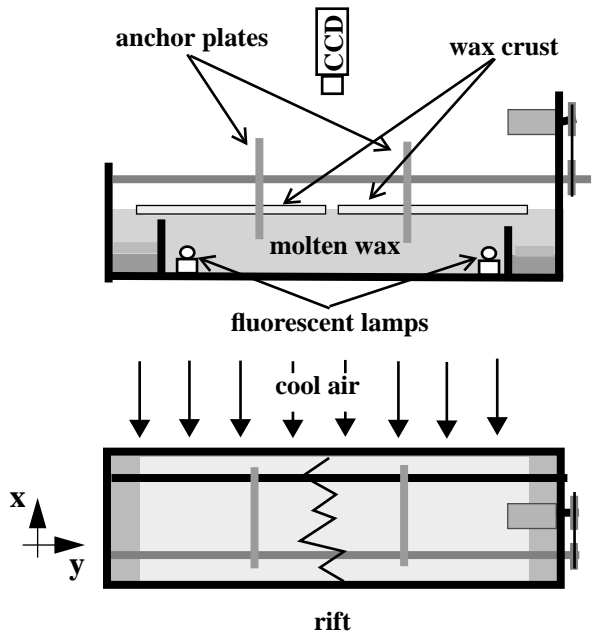

Figure 1: Experimental setup schematically: side and top view.

with stainless steel side walls and an aluminum bottom. It was filled to a height of 8 cm with Shellwax 120. The bottom was heated by a regulated circulating water bath to $(57.5 \pm 0.05)\,°C$, slightly above the $54°C$ melting temperature of Shellwax 120. To solidify the wax surface, air of temperature $(25.5 \pm 0.5)\,°C$ was blown across the long side of the apparatus [9]. Two vertical stainless steel anchor plates of 3 mm thickness were frozen into the wax. A micro-stepping motor was used to move the two anchor plates apart, each at the same speed, ranging from $1\mu m/sec$ to $1\,mm/sec$. To avoid a buildup of solid wax at the ends of the tray, each wax crust was melted off in warmer regions of lateral size $10\text{x}20\,cm^2$. Each region was heated by an electric film heater mounted on an aluminum plate of size $100 \times 200 \times 3\,mm^3$ and insulated from the temperature controlled bottom by a layer 3 cm thick polyurethane foam. To minimize thermal cross-talk, two 6 cm high stainless steel plates separated the warmer regions from the main part of the tray, allowing the solid crust to pass over. The wax layer was illuminated from below by two fluorescent lamps placed into the molten wax and extending over the width of the tray, allowing the rift to be visualized from above with a CCD camera connected to a digital image processing system. Due to the light diffusing properties of the rigid wax, the crust appeared white, while closer to the rift the ductile wax was darker, and regions of molten wax appeared black.

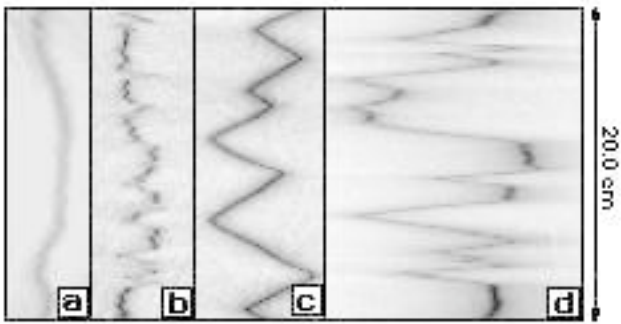

Figure 2: Rift patterns at different spreading rates (a) $v = 35$ $\mu m/sec$, $t = 4800\,sec$; (b) $v = 85\,\mu m/sec$, $t = 1900\,sec$; (c) $v = 177\,\mu m/sec$, $t = 1400\,sec$; (d) $v = 709\,\mu m/sec$, $t = 360\,sec$. The arrow indicates the scale of the pattern.

In a typical run, first the cooling air was turned off and all the wax was melted. The anchor plates were positioned about 5 cm apart and the cooling air was turned on again. After about 10 minutes the wax had solidified to a rigid crust approximately 2 mm thick, and a straight rift was cut with a sharp knife parallel to the anchor plates [10]. At $t = 0$ the micro-stepping motor was turned on and moved the two wax plates apart with constant velocity. The newly created crust had a thickness less than 2 mm [11]. During the run digital images of the rift where taken at constant time intervals. A run ended when the anchor plates reached the warmer regions.

Rift patterns obtained at four different spreading rates are shown in Fig 2. The images were taken after a time t when initial transients had died out and clearly illustrate the existence of different rift formation regimes.

For plate pulling speeds (or, equivalently, spreading rates), $v \leq 40\,\mu m/sec$, a straight rift remained flat and frozen over as shown in Fig. 2a. For $40 < v \leq 120\,\mu m/sec$, a spiky rift pattern developed. The rift was dominated by oblique rift segments (Fig. 2b), which for increased spreading rate were almost parallel to the spreading direction. The oblique rift segments were frozen over and showed fracture and rift slippage. These regions were hardly visible in a linearly scaled gray scale image and a false colormap was used in Fig. 3a to enhance the rift structure. Thick frozen wax corresponds to blue, thinner frozen wax is green/yellow, and upwelling hot liquid wax is red. It is remarkable that the rift is torn only in a few places where the rift segments are almost perpendicular to the spreading direction.

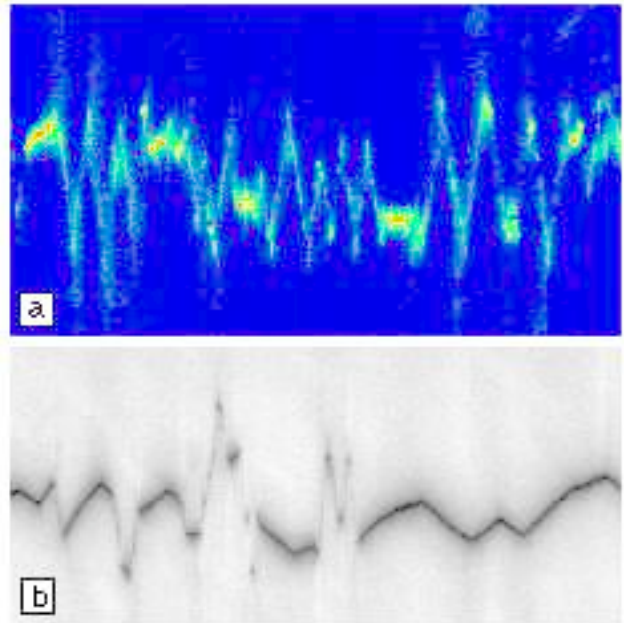

Figure 3: (a) Rift structure in false color for $(v = 107\,\mu m/sec, t = 2400\,sec)$. Torn regions of molten upwelling wax are red, while crystallized wax is blue. (b) Rift structure in the coexistence region ($v = 142\,\mu m/sec$, $t = 2250\,sec$). The width of the images corresponds to 20 cm.

In the range of $120 < v \leq 170\,\mu m/sec$, the spiky pattern coexisted with a regular zig-zag pattern (Fig. 3b), which was the selected rift structure for $v > 170\,\mu m/sec$, as shown in Figs. 2c and 4.

In the zig-zag region we did not observe the partial freezing of the rift that characterized the spiky state. Instead, the rift was open along its whole length, as evidenced by the uniformly black color of the rift in Figs. 2c and 4. Two solidification fronts propagating from each crust solidified the upwelling wax. Contrary to the spiky regime, where stress and temperature field must be important factors for the rift formation, here the two fronts can interact only via the temperature field. Upwelling hot wax, latent heat release due to solidification at each edge of the rift [12], and cooling air temperature are the obvious factors determining the width of the rift.

With increasing spreading rate the oblique rift segments became steeper and steeper, finally forming a pattern of fault-like slips parallel to spreading direction separated by straight regions perpendicular to it (Fig. 2d). This latter pattern is quite similar to the one observed and interpreted earlier as transform faults [1-3,5].

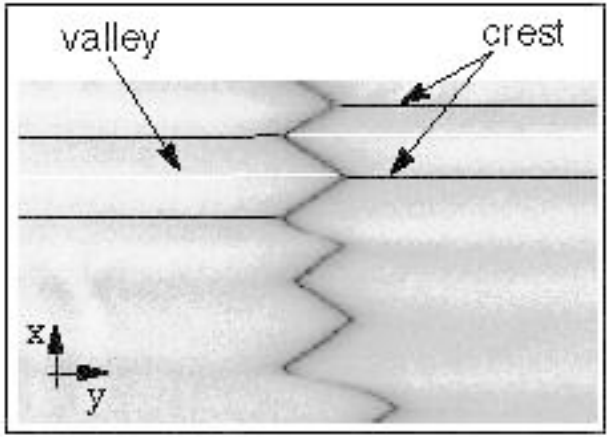

Figure 4: Zig-zag pattern ($v$ = 241 µm/sec, t = 1200 sec, width = 8.7cm) showing the three dimensional structure behind the rift. Liquid wax is black. Crystallized solid wax is alternating between dark and bright (for details see text).

The solid crust behind the rift had a topography of crests and valleys as shown in Fig. 4. To visualize the crustal topography, the fluorescent lamps were angled slightly of the $x$-axis adding illumination from the side. Regions tilted towards the high intensity side scatter more light and appear brighter than regions tilted away from the high intensity side of the tray. The crustal topography may be explained by the fact that the crust along any line parallel to the $x$-axis varied in age, $i.\ e.,$ the crust further from the rift was created earlier and had cooled for a longer time, leading to an increased thickness. While cooling, the older crust shrunk and stresses developed pulling the younger regions apart. As a result a straight line of ductile wax (crest) was left behind the pointed side of the apices. The crests were elevated by about 1 mm relative to the bottom of valleys built by the oldest wax. This topography can be understood by the shrinkage of the crystalline wax at the surface while the wax below was still in its ductile phase.

In order to systematically characterize the rift patterns, we measured the local angle between the rift and the $x$-axis (see Fig. 1). From the digital images the position of the rift $(x_i, y_i)$ was found by locating the darkest pixel along each row parallel to the $y$-axis. The obtained set of points $y_i(x_i)$ were smoothed three times with a three-point running average in order to minimize effects due to finite CCD pixel resolution. We then calculated the slope with a fourth-order finite difference formula and from it the characteristic angle $\varphi_i(x_i)$. The angles $\varphi_i$ were plotted in a histogram, where each bin count was weighted by the length of each segment [13]. The maximum of the histogram was chosen to be the characteristic rift angle $\varphi$. Fig. 5 shows the characteristic angles obtained for different pulling speeds. The plot shows both the selected angles of the spiky regime at slow spreading rates and of the zig-zag regime at faster spreading rates. While the angle in the spiky regime rapidly approaches the limit of 90° causing fracture and rift slippage, the angle in the zig-zag regime increases more gradually.

In the zig-zag regime we observed that the rift width was approximately constant up to spreading rates of about 500 µm/sec, $i.\ e.$, the growth velocity $v_g$ of each solidification front perpendicular to the edges of the wax plates was constant. Under these conditions the rift can remain perpendicular to the pulling direction for spreading rates $v \leq v_g$, while for $v > v_g$ the constant rift-width cannot be maintained. In other words, for the rift to show a constant rift width the solidification fronts have to grow at an angle to the $x$-axis causing an oblique rift. This characteristic angle can be easily obtained from trigonometry: $\varphi = \mathrm{acos}(v_g/v)$. Fits of the experimental data to the geometrical model give excellent agreement. As shown in Fig. 5, the experimental data may also support two regimes for low and high spreading rates [14], suggesting a more complex pattern-formation mechanism. In particular, stresses generated by wax shrinkage, which lead to the deformation of the crust are not included.

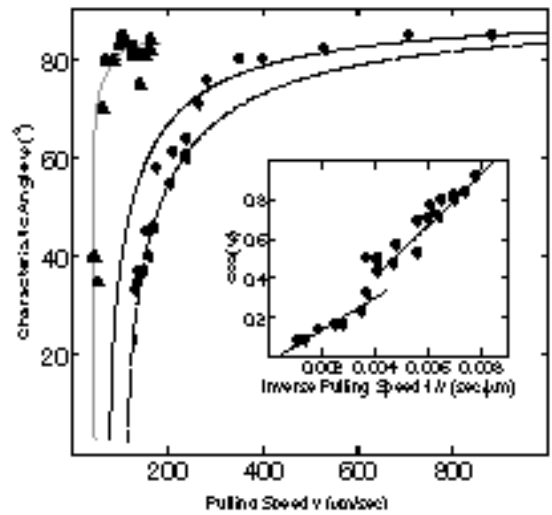

Figure 5: Characteristic angle versus spreading rate: spiky - regime (triangles); zig-zag regime (dots). Also shown are the fits to the geometrical model for $120 \leq v \leq 266$ µm/sec (solid line), and $v > 266$ µm/sec (dashed line). The grey line is a guide to the eyes.

So far only static properties of the rift formation have been discussed. Due to the complexity of the pattern evolution in spiky regime we limit our discussion to the zig-zag regime. As shown in Fig. 6 a well developed zig-zag pattern coarsened by decreasing the number apices in steps of two. The coarsening process was always initiated at an apex of the rift pattern, where the apex developed into a straight front parallel to the x-axis which propagated with a velocity equal to or less than 1/4 of the spreading rate. It moved under constant deceleration so as to create a new apex on the opposite side and to eliminate the two adjacent apices. Sometimes we observed multiple coarsening processes (as visible in Fig. 6), but due to the finite length of the experiment we could not capture the late stages of the coarsening process. The coarsening itself is only explainable by an asymmetrical growth process that allows the rift edge on one side to grow faster than the opposite one. This behavior is totally unaccounted for in the geometrical model and we believe that the nonlocal stress field due to the shrinkage of the wax is an important factor in the coarsening process. In particular, the wavy topology behind

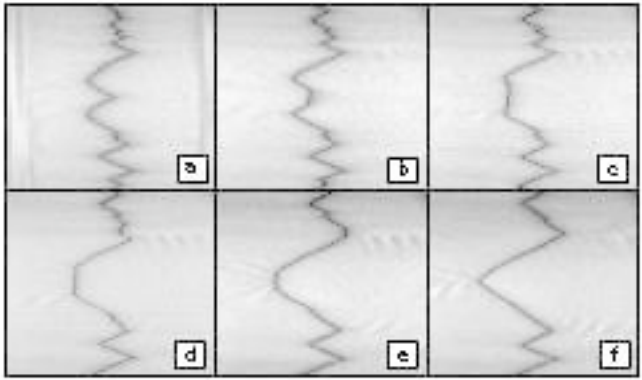

Figure 6: Coarsening of rift pattern at $v = 213$ μm/sec. The images in the sequence are taken 210 sec apart and each have a width of 20 cm.

the apices in Fig 6 may be contributed to stresses in the crust.

In conclusion, we have shown that the rift pattern phase diagram is surprisingly complex. For the zig-zag regime the geometrical model leads directly to the steep angles needed for the development of transform faults. Although the pattern morphologies obtained in our experiment are different from those of the ocean floors, the system presented above offers the possibility of testing models for rift formation. From the pattern-formation point of view, the geometrical model describing the zig-zag pattern should be universal. Indeed, the same model was found to explain triangular disclination lines in an experiment of directional solidification of nematic liquid crystals [15]. In another experiment on magnetic domains in amorphous iron films, similar zig-zag pattern were observed [16], and we believe that in this case the zig-zag structure may be triggered by a competition of the imposed velocity, given by the switch-on time of the magnetic field and the intrinsic propagation velocity of the domain walls.

EB thanks B. Shaw for bringing this problem to his attention. We have benefitted from discussions with G. Barkema, M. Grant, J. Guckenheimer, T. Molteno, S. Morris, K. Satyanarayan, D. Turcotte, and S. Watanabe, as well as from the contributions by the technical personnel of the Laboratory for Atomic and Solid State Physics. Shell Development Co. graciously supplied paraffin wax samples. We also thank C. Franck, U. Happek and A. Sievers for providing equipment in the initial stages of the experiment. This work was supported be the Alfred P. Sloan Foundation and the National Science Foundation under Contract No. DMR-9121654. JLF acknowledges support from undergraduate program of the Material Science Center at Cornell University.

* Presently at Harvard University, Cambridge, MA.
** Electronic address: eb22@cornell.edu